\documentclass[12pt]{iopart}
\usepackage{iopams,amssymb,graphicx,amsbsy,hyperref,color,CJK}

\begin{document}
\begin{CJK*}{UTF8}{mj}
\title{How modular structure determines operational resilience of power grids}

\author{Heetae Kim (김희태)}
\address{Department of Energy Technology, Korea Institute of Energy Technology, Naju 58322, Korea}
\address{Data Science Institute, Faculty of Engineering, Universidad del Desarrollo, Santiago 7610658, Chile}
\ead{kimheetae@gmail.com}
\end{CJK*}
\vspace{10pt}

\begin{abstract} 

The synchronization stability has been analyzed as one of the important dynamical characteristics of power grids.
In this study, we bring the operational perspective to the synchronization stability analysis by counting not only full but also partial synchronization between nodes.
To do so, we introduce two distinct measures that estimate the operational resilience of power-grid nodes: functional secureness centrality and functional robustness centrality.
We demonstrate the practical applicability of the measures in a model network motif and an IEEE test power grid.
As a case study of German power grid, we reveal that the modular structure of a power grid and particular unidirectional current flow determine the distribution of the operational resilience of power-grid nodes.
Reproducing our finding on clustered benchmark networks, we validate the modular effect on power grid stability and confirm that our measures can be the insightful tools to understand the power grids' synchronization dynamics.

\end{abstract}

%
%
%
%
%

\section{\label{sec:introduction}Introduction}
Modern power grids are getting more complicated to meet the increasing demand for electricity~\cite{worldbank}. 
For instance, consumers not only consume electricity but also produce it participating in a power system as prosumers.
The capacity of renewable power generation also increases, which makes power systems decentralized.
However, the distributed renewable power sources encounter the frequent power fluctuation that is hardly predictable. 
Consequently, the complex power system possesses the risk of cascading power failures that can spread through the entire power grid resulting a massive blackout~\cite{Buldyrev:2010ej,Bashan:2013ts,Yang:2017vp,Nesti:2020jg}.

To overcome the crisis, power system studies have investigated how the topological properties of power grids affect the dynamical stability. 
For example, the nodes at dangling-ends or at detour paths are known to undermine the synchronization stability of power grids~\cite{Menck:2014fn,Schultz2014detours}.
A specific set of vulnerable nodes being distinguished by their large {\it k}-core more likely trigger primary power failures~\cite{Yang:2017io}.
The topological formation of a power grid also affects its stability because the power-grid nodes interact through the connection structure~\cite{Rohden:2014fy,Kim:2016kd,Nitzbon:2017fo,Odor:2018kf,Montanari:2020cp}.
The amount of power input as well as its spatial distribution influence over the synchronization stability of power grids, which gives an insight to design decentralized power systems~\cite{Rohden:2012gl,Lee:2017dz,Kim:2019cq}.
The abovementioned studies impart effective clues to predict the synchronization stability of power grids, mainly based on the network properties of individual nodes or links.

In this study, we expand our knowledge of understanding power-grid stability further to mesosacle characteristics: the modular structure of a power grid.
The modular structure refers to the node clustering in a network formed by dense connections within a group of nodes and sparse connections between groups~\cite{Newman:2004ep,Newman:2006iq}. 
In network literatures, a modular group is called a {\it community} (we interchangeably use both in this paper), and numerous detection techniques identify communities in a given network~\cite{Newman:2004ep,Newman:2004jh,Reichardt:2004ea,Clauset:2004dz,Rosvall:2008fi,Blondel:2008do}.
It is known that the modular structure plays an important role for the function of networks.
For instance, the community structure of neuron firing in a brain network explains its neuronal function~\cite{Garcia:gza,AvenaKoenigsberger:2018ja}.
The community configuration of the three-dimensional molecular structure of human chromosome reveals how physically distant DNA loci can be functionally associated~\cite{Lee:2019ui}.
More relevantly to our study, the synchronization stability of power grids is related to the modular connection of transmission lines~\cite{Kim:2015kg}.

Indeed, the nature of power grids features modular connectivity since its installation depends on the population distribution~\cite{Nesti:2020jg}. 
Transmission lines densely cover urban areas while only a few lines span between cities.
When the modular connectivity accompanies the power-law distributed electricity consumption, extreme blackouts can break out~\cite{Nesti:2020jg}.
Suffering a massive blackout, sometimes a power grid can be partitioned into several sub-grids having a subset of the power grid still functioning, called power-grid islanding~\cite{Hamad:2011he,Rikvold:2012jn,Mureddu:2016dw}.
Despite such profound connection between the modular structure and its stability, little is known about how the modular grid structure affects functional (or operational) stability of power grids, which we will investigate in this study.

To examine the operational stability of power grids, we introduce novel measures---functional secureness centrality and functional robustness centrality---to estimate the stability of not only full but also partial synchronization recovery.
In the remainder of this paper, we introduce our measures and show that the modular structure and the power flow affect the functional stability and resistance of power grids.
First, in Sec.~\ref{sec:methods} we bring a numerical method to simulate the synchronization dynamics of power grids, and subsequently, the introduction to our proposed measures follows.
In Sec.~\ref{sec:forwhat}, we show how functional secureness centrality and functional robustness centrality can reveal the dynamical characteristics of power-grid nodes in comparison to the popular synchronization stability measure, basin stability~\cite{Menck:2013bk}, by using a six-node motif and an IEEE 24-bus model.
The modular effect on the synchronization stability is then investigated with German power grid as a case study, for which we reproduce our findings with a synthetic random model.
We summarize our findings and add our final remark at the end in Sec.~\ref{sec:discussion}.

\section{\label{sec:methods}Methods}

\subsection{\label{sec:sync}Operational resilience}
When a perturbation disturbs the synchronous state between power-grid nodes, the grid may lose its full synchrony, however, part of the nodes can still be synchronized~\cite{Menck:2014fn}.
The power grid will not be completely dead upon the partial synchronization.
Although it is not a full synchronization between all power-grid nodes, if some clustered nodes in a modular power grid remain synchronized between them to the rated frequency, the cluster can still be operational as a fully functioning power-grid island~\cite{Nesti:2020jg}.

\begin{figure*}[h]
  \hfill\includegraphics[width=0.9\textwidth]{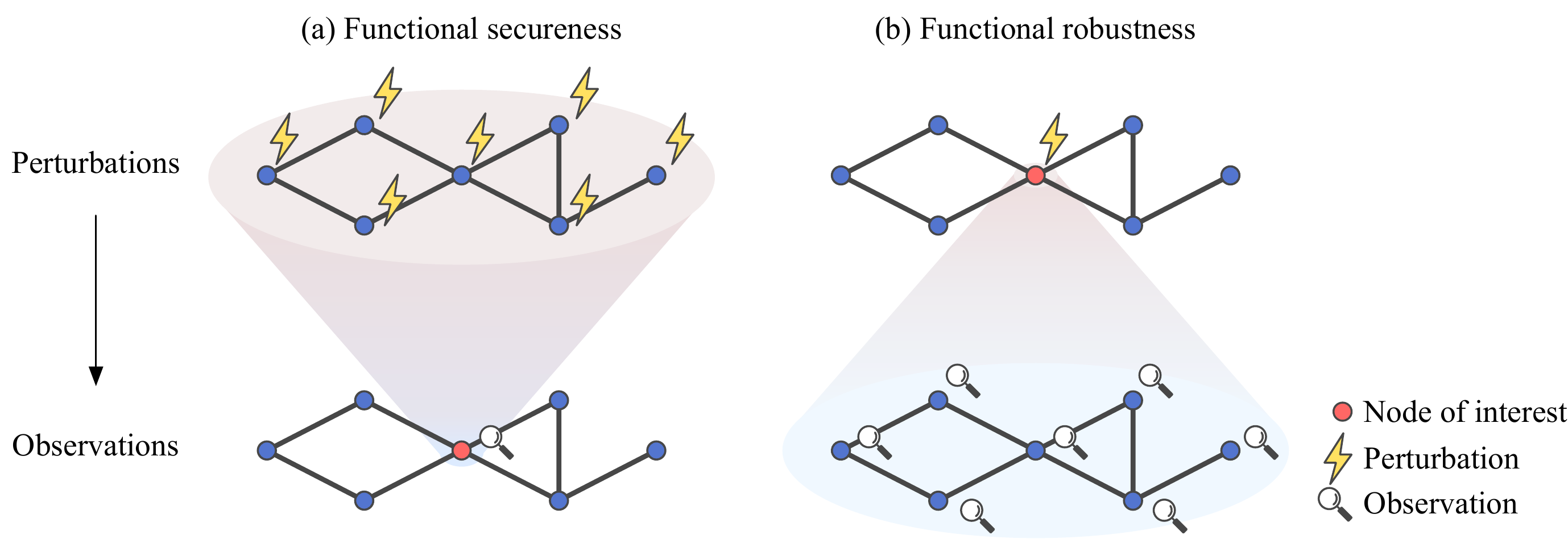}
  \caption{A schematic diagram of (a) functional secureness and (b) functional robustness of a power grid when perturbing a single node at a time. The single-node functional secureness centrality of a node measures the passive vulnerability to the node against external perturbations (to each node in turn), while the  single-node functional robustness centrality estimates how a node influential is to the grid. Each measure reflects the opposite aspect of the interaction. 
  }
  \label{fig_1}
  \end{figure*}

To estimate the operational resilience of each power-grid node, we come up with two different aspects of the influence due to perturbations: as a passive receiver and an active impacter.
In the receiver's point of view, we can measure how operationally secure a node is against perturbations to the power grid in general.
Specifically, we estimate the probability of the rated-frequency recovery of each node when perturbing the others (including itself).
We focus on the status of the node of interest recovered the rated frequency, not the entire power grid.
As it represents how secure each node's operational function---supplying electricity at the rated frequency---is, we call it functional secureness centrality (FS).
One can consider that the nodal FS is to observe the influence of the perturbations from the power-grid nodes to a node [Fig.~\ref{fig_1}(a)].

Contrariwise, it is also of interest to know how influential a node is in terms of the perturbations' focal point.
A highly influential node will break most nodes' synchronization in the power grid.
In other words, the more influential a node is, the lower functional robustness the node has. 
In this regard, we evaluate functional robustness centrality (FR) of a node from the probability of the power grid's synchronization recovery when perturbing the node.
Through this concept we can see the power-grid's response to the targeted perturbations [Fig.~\ref{fig_1}(b)].

We emphasize again that for both FS and FR, we do not limit ourselves to count only full synchronization but also consider partial synchronization as we are interested in the operational characteristics of power grids.
Note that the strict version of FR that takes into account only full synchronization corresponds to `basin stability (BS)'~\cite{Menck:2013bk}.
In this paper we call `operational resilience' when we mention both `functional secureness' and `functional robustness' and use `functional' or `operational' to describe nodes' status being synchronized to the rated frequency.

\subsection{\label{sec:measure}Synchronization dynamics of power-grid nodes}

To investigate the synchronization dynamics of power grids, we numerically analyze the interaction between power-grid oscillators.
Consider a power-grid network of $N$ oscillators (nodes) and $L$ transmission lines (links). 
Depending on how aggregate the physical power grid to a network structure, each $i$th node can represent a single bus or aggregated power-grid components within a region~\cite{Nishikawa:2015gl,Kim:2018ie}. 
The net power input $P_i$ of each node defines its role in the power grid: $P_i>0$ as a net power producer; $P_i<0$, a net power consumer; $P_i=0$, out of service due to maintenance or a branch point such as a tower.

The dynamical interaction between power-grid nodes is governed by the second-order Kuramoto type model called Swing equation~\cite{Menck:2014fn,Bergen:1981hj,Dorfler:2013ew}: 
\begin{equation}
\ddot{\theta}_i=\dot{\omega}_i=P_i-\alpha_i\dot{\theta}_i-\sum^N_{j=1}K_{ij}\sin(\theta_i-\theta_j),
\label{Swing}
\end{equation}
where $\theta_i$ is the phase deviation of node $i$ from a reference frame that is rotating at the rated frequency (50 Hz or 60 Hz), and $\dot{\theta}_i=\omega_i$ is the deviation of the angular frequency of $i$ to the reference frame. 
Each node is supposed to maintain $\omega_i=0$ to supply quality AC at the desired frequency, and a power grid is fully synchronized when $\omega_i=0$ for all nodes. 
The coupling strength $K_{ij}=K_{ji}$ corresponds to the transmission capacity between $i$ and $j$, otherwise $K_{ij}=0$ when nodes $i$ and $j$ are not connected. 

Given the initial state $\mathbf{x}=(x_1, x_2, \dots, x_N)$ with $x_i=(\omega_i,\theta_i)$ for each node, we numerically simulate the non-linear dynamical interaction between the nodes. 
In this study we use the fourth-order Runge-Kutta method~\cite{10.5555/1403886} with $dt=0.001$ for the numerical implementation of the Swing equation.
Since a small disturbance can be recovered by various local control mechanisms~\cite{Taher:2019gc,Tumash:2019fk}, we set the frequency tolerance threshold as plus or minus $\epsilon=5\times 10^{-2}$ Hz, and we consider a node $i$ is in normal operation when $|\omega_i|<\epsilon$. 

\subsection{\label{sec:single_node}Single-node FS and FR}

We estimate FS or FR of power-grid nodes by using numerical simulations with Monte Carlo method given external perturbations.
In reality, the power failures can occur to both an individual node or a group of nodes.
Either way, the perturbations can be simulated by various number of initially perturbed nodes in the power grid, and each case has different context of power failures.
For instance, the perturbations to a single node allow us to measure the FS in response to a local disturbance, which we call \emph{single-node FS}.
Besides, applying simultaneous perturbations to multiple nodes enables us to estimate \emph{multi-node FS} that represents the nodal FS against rather global disturbances such as a regional disaster affecting several nodes at the same time, an intended attack to a particular set of nodes, or a random disturbance to multiple nodes.
We first present the numerical procedure to estimate single-node FS below.

First, let the nodes in a power grid be fully synchronized at the rated frequency as $\omega_i=\omega_i^*=0$ and $\theta_i=\theta_i^*$ for all $i$.
The synchronized state $(\omega_i^*, \theta_i^*)$ is numerically acquired by setting both $\omega_i$ and $\theta_i$ as zero for all nodes followed by a long time iteration until the system converges to the synchronized state.
Then, we give a perturbation to a target node $i$ by setting $x_i(0)=(\omega_i(0),\theta_i(0))$ from a two-dimensional state space regarding $i$ with the density $\rho(x_i)$ of the state, where $\int_{x_i}\rho(x_i)d{x_i}=1$.
In this study, we also consider large perturbations and uniform probability distribution of the disturbances, so that we set $\theta_i(0)\in[-\pi,\pi)$ and $\omega_i\in[-100,100]$ uniformly at random. 
To investigate the comprehensive profile of the influence between nodes, we perturb all nodes in turn for the number $E$ of initial perturbation ensembles.

Counting the number $F^1(i,j)$ of perturbations that the node $j$ converges to $|\omega_j|<\epsilon$ when initially perturbing node $i$, we get the ensemble average $\bar{F}^1(i,j)=F^1(i,j)/E$ that represents the probability of node $j$ to recover its nominal frequency against the disturbance on $i$. 
The estimated $\bar{F}^1(i,j)$ reveals the pair-wise influence between perturbing and responding nodes, therefore, it discloses the detailed system's response to particular external impacts. 
The superscript $1$ of $\bar{F}^1(i,j)$ here indicates that we perturb a single node at each time.

Finally, we measure the single-node FS of the node $j$ as the average over all perturbed nodes:
\begin{equation}
\langle S_F^1(j) \rangle=\frac{1}{N}\sum_i{\bar{F}^1(i,j)}
\label{FS}
\end{equation}
that effectively reveals a node's operational passive influence against any random local perturbation.
For the network's perspective, the mean single-node FS, $\langle S_F^1 \rangle=\frac{1}{N}\sum_j{\langle S_F^1(j) \rangle}$ represents the network's average functional secureness.

The single-node FS of a node shows how affected the node is by the others. 
However, it is also useful to know how much a node influences to the others---the single-node FR---that can be calculated by averaging $\bar{F}^1(i,j)$ according to $j$.
Formally, the single-node FR of $i$ is then
\begin{equation}
\langle R_F^1(i) \rangle=\frac{1}{N}\sum_j{\bar{F}^1(i,j)},
\label{FR1}
\end{equation}
and naturally the mean single-node FR is identical to the mean single-node FS, $\langle R_F^1 \rangle=\frac{1}{N}\sum_j{\langle R_F^1(j) \rangle} =\langle S_F^1 \rangle$.

Alternatively, one can measure the single-node FR in a similar manner of single-node BS~\cite{Menck:2014fn,Mitra:2017jp}.
First, for each perturbation we observe the angular frequency of each node and calculate the level of synchronization by
\begin{equation}
\bar{R}^1(i,e)=\frac{1}{N}\sum_j{\mathbf{1}_{[0,\epsilon)}(|\omega_j^{i,e}|)},
\label{FR2}
\end{equation}
where the indicator function $\mathbf{1}_{[0,\epsilon)}(|\omega_j^{i,e}|)$ yields $1$ if $|\omega_j^{i,e}|<\epsilon$ and $0$ otherwise, and $\omega_j^{i,e}$ is the angular frequency deviation of node $j$ at $e$th realization when perturbing a node $i$. 
By averaging $\bar{R}^1(i,e)$ over the ensembles we get $\langle R_F^1(i) \rangle=\frac{1}{E}\sum_e{\bar{R}^1(i,e)}$.
In other words, the single-node FR means the average proportion of the system's nodes that return to their desired state after a particular node is perturbed.
The single-node BS, however, counts only fully synchronized cases ($\bar{R}^1(i,e)=1$) over the ensembles:
\begin{equation}
\langle{S}^1_B(i)\rangle=\frac{1}{E}\sum_e{\mathbf{H}(1-\mathbf{1}_{[0,\epsilon)}(|\omega_j^{i,e}|))},
\label{BS}
\end{equation}
where $\mathbf{H}(\cdot)$ is the Heaviside step function.
By definition, single-node FR of a node is greater than or equal to single-node BS of the node, $\langle{R}^1_F(i)\rangle\geq\langle{S}^1_B(i)\rangle$.

Measuring the single-node FS and FR, we are interested in the interplay between the target node and the system, which consists of all nodes including the target node itself, to observe the system-wide reaction. 
In the case studies that we are showing in the following section, the effect of the self-interaction (when $i=j$) is not significant and does not mask the influence of the other nodes.
Indeed, the self-influence will be negligible in the typical size of power grids that have hundreds of nodes.
Instead, considering the self-interaction enables us to interpret the results along with the conventional synchronization stability measure, basin stability, in which the self-interaction is also included. 
Therefore, we consider the corresponding terms in the summation. 
However, of course, it is also possible to strictly exclude the self-interaction by removing them when it is necessary.

\subsection{\label{sec:multi_node}Multi-node FS and FR}

In various circumstances, an external disturbance can affect several nodes simultaneously.
In other words, it is necessary to investigate the operational resilience upon the perturbations on multiple nodes.
One can effectively estimate the impact of the multiple disturbances by using the number $m$ of initially perturbing nodes~\cite{Mitra:2017jp}.
The numerical procedures of measuring the multi-node FS and multi-node FR are similar to those of single-node cases, but it requires multiple nodes to be initially perturbed as described in the following.

Instead of perturbing a single node $i$, we initially perturb $m\geq 1$ node(s) selected uniformly at random at every realization $e$.
We denote the perturbed $m$-node set as $M^m_e$ that can change at every realization.
Optionally, one can consider the failure probability of each node based on the real failure frequency instead of the random selection or use a specific fixed target node set $M_{uni}$, which we will discuss soon.
As the single-node estimations, we assign $\theta_i(0)\in[-\pi,\pi)$ and $\omega_i=\in[-100,100]$ uniformly at random for a node $i\in M^m_e$ or $\theta_i(0)=\theta_i^*$ and $\omega_i(0)=\omega_i^*$ otherwise.
For each node $j$ we count the number $F^m(j)$ of realizations that the node $j$ converges with the criteria $\epsilon$.
The average over total $E$ realizations yields the probability of node $j$'s nominal frequency recovery to the $m$-node disturbances, the $m$-node FS of $j$:
\begin{equation}
\langle S_F^m(j)\rangle=\frac{F^m(j)}{E}.
\label{mFSi}    
\end{equation}
Accordingly, we define the mean multi-node FS as $\langle S_F^m\rangle=\frac{1}{N}\sum_j\langle{S_F^m(j)}\rangle$, which represents the power grid's averaged functional secureness for $m$ simultaneous perturbations.
Note that we get the mean multi-node BS regarding the $m$ random initial perturbations as
\begin{equation}
  \langle{S}^m_B\rangle=\frac{1}{E}\sum_e{\mathbf{H}(1-\mathbf{1}_{[0,\epsilon)}(|\omega_j^{{M^m_e},e}|))},
  \label{multi_BS}
  \end{equation}
which is originally suggested by Mitra {\it et al.}~\cite{Mitra:2017jp}.

When we randomly perturb multiple nodes, multi-node FS represents the functional secureness according to the perturbation size in general.
However, one may want to analyze the grid's response about specific target nodes.
The influence of the targeted multi-node disturbance can be measured by using the unique $m$-node set $M_{uni}$ to be attacked.
Through $E$ times of Monte Carlo realization, we estimate the multi-node FS corresponding to the specific node set $M_{uni}$ as:
\begin{equation}
  \langle {S}_F^{M_{uni}}(j) \rangle =F^m(M_{uni},j)/E,  
\label{mFSu}
\end{equation}
where $F^m(M_{uni},j)$ is the number of synchronization recovery of node $j$ over the simultaneous perturbations on $M_{uni}$.
The average over $j$ represents the overall influence of the $m$-node set to the power grid, which corresponds to the multi-node FR of $M_{uni}$:
\begin{equation}
\langle {R}_F^m(M_{uni}) \rangle = \frac{1}{N}\sum_j{\langle {S}_F^{M_{uni}}(j) \rangle}.
\label{mFS}
\end{equation}

An application of the mean multi-node FS is to diagnose a power grid's security limit, $m_{crit}$.
The security limit refers to the critical number of failure nodes, where its functionality decreases to lower than a minimum threshold $\langle{S_F}\rangle_{th}$~\cite{Mitra:2017jp}.
Given a threshold $\langle{S_F}\rangle_{th}$, we measure $\langle S_F^m\rangle$ by increasing $m$ from $1$ to $N$, and
we find $m_{crit}$ where $\langle{S_F}\rangle_{th} \geq \langle S_F^m\rangle$.
The resulting security limit represents the maximum number of simultaneous nodal perturbation that the power grid can guarantee certain functionality. 
It can also be interpreted together with the critical number of multi-node BS~\cite{Mitra:2017jp} to understand the dynamical characteristics of the power grid in a more comprehensive way.

\section{\label{sec:forwhat}What operational resilience can reveal} 

Functional secureness and robustness can reveal the unique traits of the operational resilience of power-grid nodes. 
In this section, we demonstrate the usefulness and practicality of our measures by comparing them to the popular full-synchronization measure, basin stability, on a small motif and an IEEE test grid.

\subsection{\label{sec:6-nodes}Single-node operational resilience} 

\begin{figure*}[h]
\hfill\includegraphics[width=0.9\textwidth]{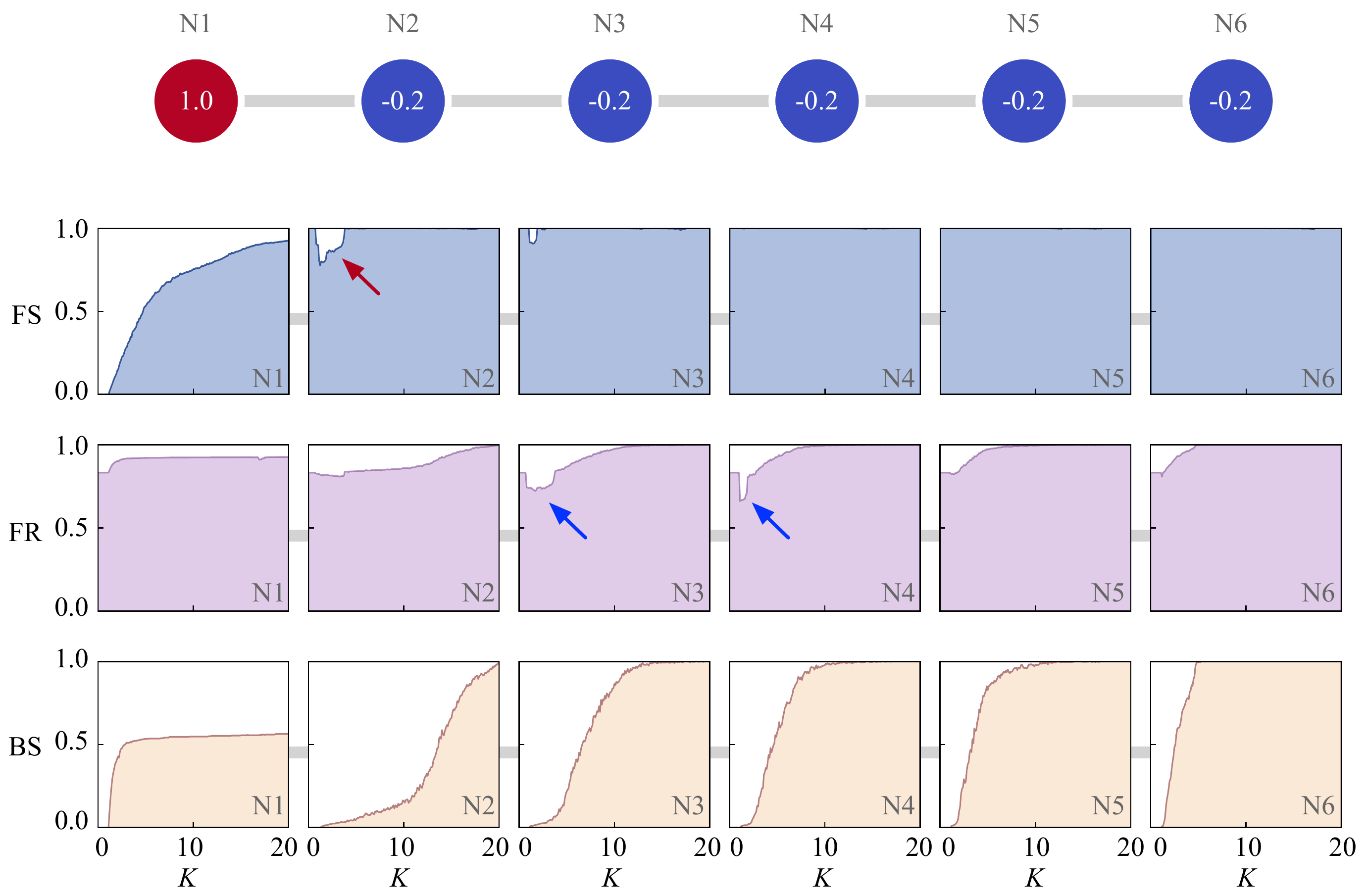}
\caption{The single-node FS, FR, and BS of the nodes in a six-node motif. The motif consists of a power producer ($P_{N1}=1$) and consumers ($P_i=-1/5$ for $i\in\{N2, N3, N4, N5,N6\}$). All links have the same $K_{ij}=K$ for $0\leq K\leq 20$. The transition patterns of FS, FR, and BS reveal different aspects of the power-grid nodes' synchronization characteristics.}
\label{fig_2}
\end{figure*}

To understand what each measure can reveal, we estimate FS, FR, and BS in a network motif with six nodes.
The motif is designed to analyze the effect of topological distance on the power-grid nodes' synchronization.
In the six-node chain shown in Fig.~\ref{fig_2}, a node supplies a unit amount of power and the others consume it equally ($P_{N1}=1$, $P_{i\in\{N2, N3, N4, N5, N6\}}=-0.2$) so that $\sum_i P_i=0$.
At the steady state when all nodes are synchronized, we perturb a single node at each instance of the total $E=500$ realizations. 
We repeat it for each node to measure FS, FR, and BS for $K_{ij}=K$, where $0\leq K\leq 20$ for the increment of $0.1$.
Observing the transition pattern of the three measures, we can disclose the comprehensive nature of the synchronization stability of the nodes from different perspectives~\cite{Kim:2019cq,Kim:2018do}.

The first row of Fig.~\ref{fig_2} shows the connection structure of the motif, and the second row is the single-node FS of the nodes as a function of $K$.
Recall that single-node FS represents a node's response against random perturbations to the power grid in general.
Hence, the result illustrates the nodal stability of normal operation at given $K$.
For example, the producer node (N1), which has the largest $|P_i|$, is showing the lowest single-node FS than the others at all $K$ values.
It means that the centralized and unitary power producer can more likely be out of the rated frequency and operationally unstable.

The transition shape of the single-node FS values also plays a role as the fingerprint of the operational stability of power-grid nodes.
For instance, the single-node FS of N1 monotonically increases according to $K$. 
It means that the unstable producer N1 becomes more stable with the larger $K$ value. 
It is understandable because large coupling strength enhances the interaction between nodes to release the disturbance. 
Therefore, the large input of N1, which can disturb itself as well, can be more easily canceled out by the other nodes when the interaction is sufficient. 

However, N2 does not show the monotonic increase but features a dip in a range of small $K$ values (Fig.~\ref{fig_2}, red arrow).  
When $K$ is negligible, the interaction is not strong enough to spread the disturbances from either the large $P_{N1}$ or the rebound impact of random perturbations from the grid, so that N2 can maintain its initial synchronization. 
However, when the coupling strength becomes stronger (increasing $K$), the large power input of N1 can interrupt the synchrony of N2, therefore the N2's FS decreases.
With even larger $K$, since the coupling strength is now enough to ensure the global interaction between nodes, the influence from N1 or the random perturbations to the grid is absorbed by the whole system, so that the FS of N2 increases again. 
The temporal dipping phenomenon is related to the topological distance to N1 that has large $|P_i|$, thus the size of the valley is smaller in N3 than that of N2, and it disappears in N4--6.
Considering this, the transition pattern of FS let us know the scale of $K$ where their interaction starts to affect.

The transition pattern of the BS (the bottom row in Fig.~\ref{fig_2}) also illustrates the gradual enhancement of the synchronization stability of N1 and the topological effect regarding the distance from the N1 to the other nodes. 
However, what BS represents is the full synchronization of the whole grid, whereas FS indicates the nodal operational viability based only on each of their own synchrony.
The results of FS clearly show that N4, N5, and N6 are very stable and, therefore, can form an operational island in which the nodes are constantly synchronized when externally perturbed.

As we defined such, FR indicates the influence of local perturbations to the entire power grid as BS does, but it also considers partial synchronization.
Therefore, one can interpret FR in the same way as we do with the BS, however, in the operational point of view, and it is always guaranteed that $\langle R^1_F(i) \rangle \geq \langle S^1_B(i) \rangle$ as shown in Fig.~\ref{fig_2}.
For example, the FR of each node in Fig.~\ref{fig_2} shows how robust the node is.
The power grid is most vulnerable when the perturbations hit N3 and N4 whose positions are central in the power grid (Fig.~\ref{fig_2}, blue arrows). 
The central location means that the average topological distance from those central nodes to the others is short, therefore the disturbance can spread out to the other nodes easily.
This finding supports the low synchronization stability of high-betweenness nodes~\cite{Kim:2016kd}.

It is interesting to note that the range of $K$ value where the single-node FR decreases in N3 or N4 is identical to the $K$ range where the single-node FS decreases in N2 or N3, respectively. 
It is due to the fact that the interaction between nodes at the corresponding range of $K$ becomes just enough to disturb neighbor nodes, but still not sufficient to release the stress within the entire network.
In practice, using single-node FR, one can identify the most vulnerable target in a power grid.
For instance, analyzing single-node BS, N1 seems to be most critical, since it breaks the full synchronization of the power grid most.
However, in the operational point of view, the perturbation at the dipping $K$ value of N3 or N4 can desynchronize the most nodes in the power grid.
Therefore, FR can be effectively applied to investigate the robustness of power grids.

\subsection{\label{sec:IEEE24}Multi-node operational resilience}

Single-node FS and FR are measured based on perturbations to a single node, therefore, it is useful to investigate the nodal influence (either from or to the node) in a power grid.
Besides, the consequence of multiple perturbations can be analyzed by multi-node FS and FR.

\begin{figure*}[h]
\hfill\includegraphics[width=1\textwidth]{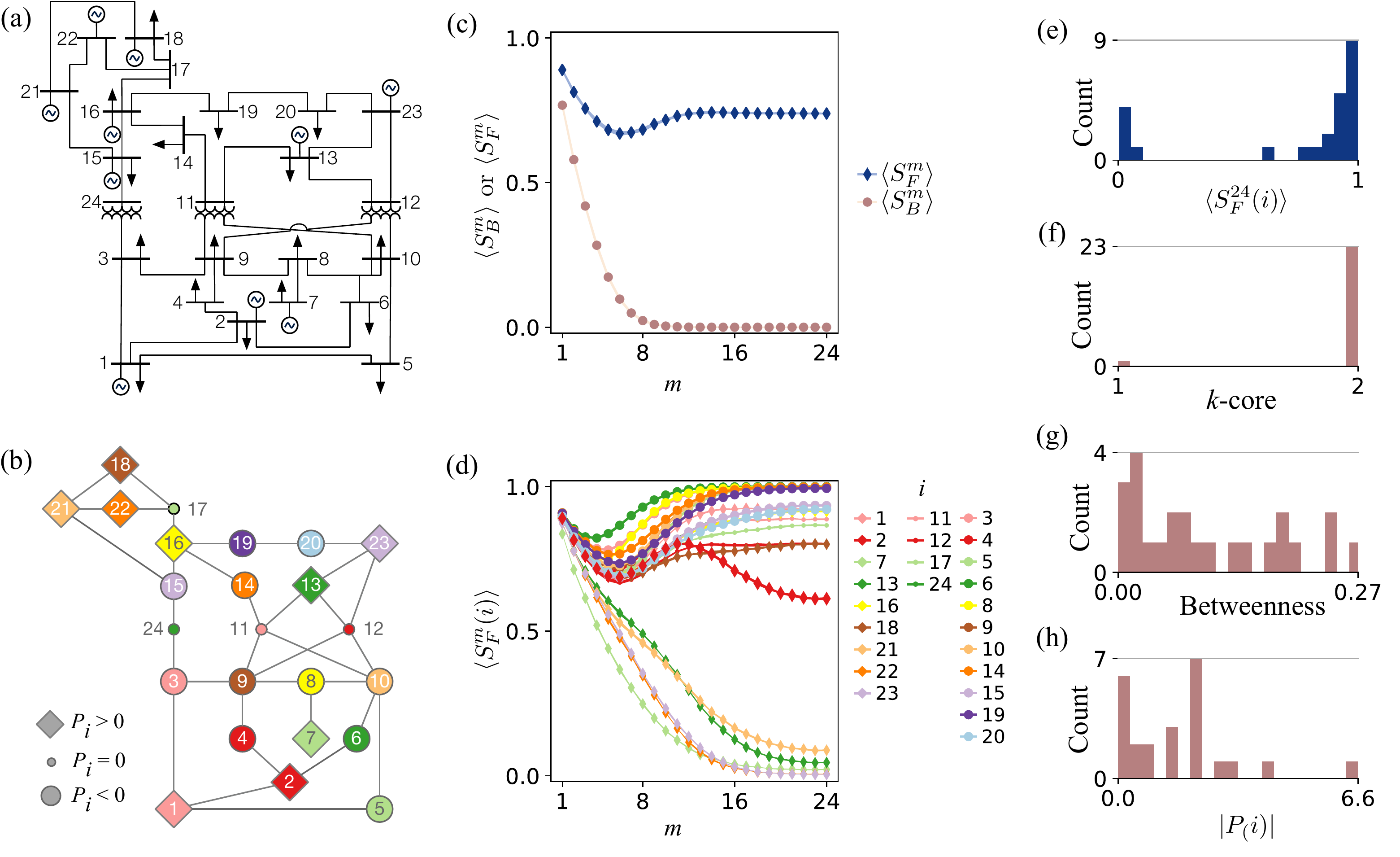}
\caption{The multi-node FS and FR analysis of IEEE test grid. (a) The IEEE test grid consists of 24 buses with 10 generators, 3 transformers, 17 loads, and 34 transmission lines. (b) We convert it to a power-grid network with 9 producers, 3 junctions, and 11 consumers. (c) The functionality of IEEE 24 test grid measured by mean multi-node FS sustains against large simultaneous perturbations, while mean multi-node BS exponentially decays. (d) The individual multi-node FS value of each node reveals that there are distinct groups of nodes according to their functional secureness upon large perturbations. For both (c) and (d), the markers represent the averaged results from 10 ensembles for 10000 realizations each. The filled-curve between markers indicate the standard deviation, which are smaller than the marker size. (e) These characteristics are only observable in multi-node FS and not discovered in other centrality measures such as (f) $k$-core, (g) shortest-path betweenness, and (h) degree.
}
\label{fig_3}
\end{figure*}

For a case study, we use an IEEE reliability test grid~\cite{Ordoudis:2016ue} that consists of 10 power generators, 17 loads, 3 transformers, and 34 transmission lines (Fig.~\ref{fig_3}a).
Considering each bus as a power-grid node and mapping the detailed characteristics of the power components such as voltage and admittance into the network attributes, we convert the test grid into a network structure (IEEE24, Fig.~\ref{fig_3}b) for the multi-perturbation analysis.
For example, each node $i$ has $P_i$ value as the net power input of the involved power components. 
Hence, in IEEE24 we have 9 net producers ($P_i>0$), 3 junctions ($P_i=0$), and 11 net consumers ($P_i<0$) as illustrated by diamonds, small circles, and large circles in Fig.~\ref{fig_3}b, respectively.
See the attributed input power of each node, $P_i$, in Table\ref{table:IEEE_node} and further detailed conversion process with attributes in Ref.~\cite{Kim:2019cq}.
To measure the multi-node statistics we selected $m$ nodes uniformly at random in each perturbation of total $E=10000$ realizations and then averaged the results from $10$ ensembles.

\begin{table}
\caption{\label{table:IEEE_node}The input power $P_i$ and the role of each node in IEEE24. The values are taken from the original data~\cite{Ordoudis:2016ue}, and the node indices correspond to the network representation shown in Fig.~\ref{fig_3}(b).}

\begin{tabular*}{\textwidth}{@{}l*{15}{@{\extracolsep{0pt plus12pt}}l}}
\br
Node & Role & $P_i$        & Node & Role     & $P_i$         \\ 
\mr
1     & Producer   & 0.2375 & 13    & Producer  & 2.77125 \\
2     & Producer   & 0.3725 & 14    & Consumer   & $-2.295$ \\
3     & Consumer & $-2.12625$ & 15    & Consumer & $-1.59625$ \\
4     & Consumer  & $-0.8775$ & 16    & Producer  & 0.36875 \\
5     & Consumer & $-0.84375$ & 17    & Junction      & 0.0 \\
6     & Consumer    & $-1.62$ & 18    & Producer  & 0.05125 \\
7     & Producer    & 2.015 & 19    & Consumer    & $-2.16$ \\
8     & Consumer   & $-2.025$ & 20    & Consumer & $-1.51875$ \\
9     & Consumer & $-2.05875$ & 21    & Producer      & 4.0 \\
10    & Consumer   & $-2.295$ & 22    & Producer      & 3.0 \\
11    & Junction      & 0.0 & 23    & Producer     & 6.60 \\
12    & Junction      & 0.0 & 24    & Junction      & 0.0 \\ 
\br
\end{tabular*}
\end{table}

The result of mean multi-node FS shows that almost 80\% of the nodes in IEEE24 can be operating regardless of the perturbation size [Fig.~\ref{fig_3}(c)]. 
Contrary to the transition pattern of mean multi-node BS that exponentially decays as a function of $m$~\cite{Mitra:2017jp}, surprisingly, the initially decreasing mean multi-node FS at small $m$ turns to increase again, and then converges.
The mean multi-node FS values reveal that the majority of the nodes in IEEE24 will still be synchronized even all 24 nodes are initially perturbed (when $m=24$).
Note again that the mean multi-node FR is identical to the mean multi-node FS by definition, therefore, we do not illustrate.

To explore the origin of the high functional secureness of IEEE24's nodes, we investigate the multi-node FS from each node's point of view.
Analyzing the multi-node FS of each node, we find that a group of nodes in IEEE24 sustain the overall functionality of IEEE24 by individually retaining the rated frequency, even though the full synchronization is broken [Fig.~\ref{fig_3}(d)].
Based on the transition pattern of multi-node FS, we can separate the nodes into two groups: one is monotonically decreasing as multi-node BS does, and another group is recovering its synchrony at large $m$. 
The characteristic segregation between the nodes upon simultaneous perturbations [Fig.~\ref{fig_3}(e) when $m=24$ for example] is a unique finding that can not be observed by other network properties, which are known to have strong correlation to the power-grid stability, such as $k$-core~\cite{Yang:2017io} [Fig.~\ref{fig_3}(f)], betweenness~\cite{Kim:2016kd} [Fig.~\ref{fig_3}(g)], and the absolute amount of input power~\cite{Kim:2019cq} [Fig.~\ref{fig_3}(h)].

It is worth to note that the monotonically decreasing group consists of the top five largest power producers (node index: 7, 13, 21, 22, and 23).
It means that the large power producers are particularly vulnerable to the global disturbances in IEEE24.
Although more rigorous analysis is required to investigate the relationship between the size of perturbation and the distinct response of the functional stability of nodes, the results clearly demonstrate that multi-node FS can reveal the detailed operational characteristics of power-grid nodes upon multiple perturbations that we can not see through multi-node BS.

\section{\label{sec:results}Results}
\subsection{\label{sec:germany}German power grid}
FS and FR can reveal the synchronization characteristics of power grids as we looked into so far.
As a case study on a real power grid, we apply them to German power grid and show the influence of the modular structure on the operational resilience. 
We use the publically opened network dataset of German power grid~\cite{Egerer2016Open} (ELMOD-DE) with detailed attributes of each node.
Specifically, it consists of 438 nodes with the information about the amount of power generation and load of the associated power facilities.
We first set the net power input $P_i$ of node $i$ as the sum of the moderate power production (40\% of its generation capacity), load, and the amount of power trade (export or import) at the node.
For the nodes that lack the detailed information, we equally assign a unit value so that the network satisfies $\sum_i P_i=0$.
The damping coefficient $\alpha_i=\alpha=0.1$ for all nodes, and the 662 transmission lines have a unit transmission capacity $K_{ij}=K=0.49$.  
To estimate single-node FS, FR, and BS, we perturb total $E=500$ times to each node.

\begin{figure*}[h]
  \hfill\includegraphics[width=1\textwidth]{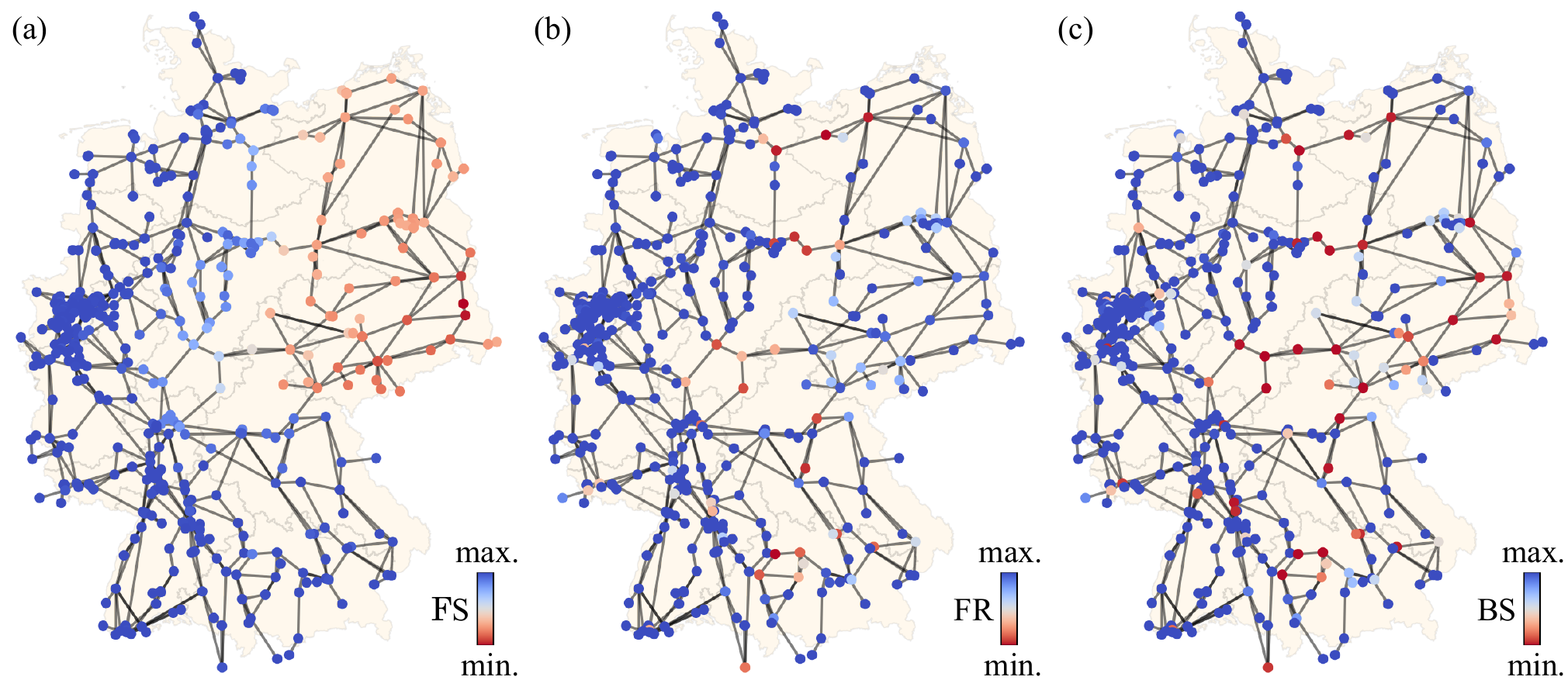}
  \caption{The single-node FS, FR, and BS of the German power-grid nodes from ELMOD-DE dataset~\cite{Egerer2016Open}. (a) The single-node FS of the nodes decrease from the west to the east resulting in a group of nodes whose single-node FS is lower than the others in the eastern region. (b) The nodes at the boundary of the group with low single-node FS show low single-node FR. (c) The distribution patterns of single-node FS and FR of German power-grid nodes do not appear in the result of single-node BS.
  Color bar scale: the node color changes linearly between minimum $0.6$ and maximum $1.0$ for FS; minimum $0.6$, maximum $1.0$ for FR; minimum $0.006$, maximum $1.0$ for BS.}
  \label{fig_4}
  \end{figure*}

Interestingly, we find that the single-node FS values of German power-grid nodes are distributed in descending order from the west to the east (Fig.~\ref{fig_4}a).
The most functionally secure nodes are observed in the western region ($\max{(\langle S^1_F(i)\rangle)}=1.0$), while the most frequently desynchronized nodes ($\min{(\langle S^1_F(i)\rangle)}=0.6$) are in the Eastern Germany.
In the central region, the single-node FS gradually decreases making a declining trend from the west to east.
The distribution pattern of the single-node FS in German power grid tells us that the Eastern (Western) Germany is more vulnerable (favorable) to sustain the rated frequency when the system is perturbed.

We also estimate which node can easily disturb the global synchronization by measuring single-node FR. 
As we already observed from the motif in Fig.~\ref{fig_2} (blue arrows), the nodes that are in central location show low single-node FR in the German power grid [Fig.~\ref{fig_4}(b)].
It is probable that the central nodes would be more influential than the other peripheral nodes for the similar reason to the six-node motif.
However, note that the nodes with low single-node FR appears at the center between the west and the east, not the north and the south, therefore, the nodes are aligned in a vertical way along the north-south direction.
The characteristic distribution patterns of the single-node FS and FR of German power-grid nodes do not appear in the single-node BS estimation [Fig.~\ref{fig_4}(c)]. 
In the following, we attempt to understand the distribution patterns of the single-node FS and FR values in relation to the community structure of the German power grid.

To analyze the community structure of the German power grid, we use Louvain community detection algorithm~\cite{Blondel:2008do} that is based on modularity optimization~\cite{Newman:2004ep,Newman:2006iq} and implemented as a Python package~\cite{community}.
Modularlity is based on the comparison between the current and random connection structures as:
\begin{equation}
  Q = \frac{1}{2m} \sum_{i \ne j} \left[ \left( A_{ij} - \gamma \frac{k_i k_j}{2m} \right) \delta(g_i,g_j) \right ] \,,
  \label{eq:modularity}
  \end{equation}
where $A_{ij}$ is the adjacency matrix that includes the connection structure of a network, $k_i$ is the number of connections (degree) of node $i$, $g_i$ is the index of node $i$'s community, $\delta$ is the Kronecker delta, and $m$ is the total number of links serving as a scaling factor.
The resolution parameter $\gamma$ decides the size of detected communities: the smaller $\gamma$, the larger a community.
In this study, we use rather a heuristic way to generate a reasonable community formation: five communities with $\gamma=0.2$.

\begin{figure*}[h]
\hfill\includegraphics[width=1\textwidth]{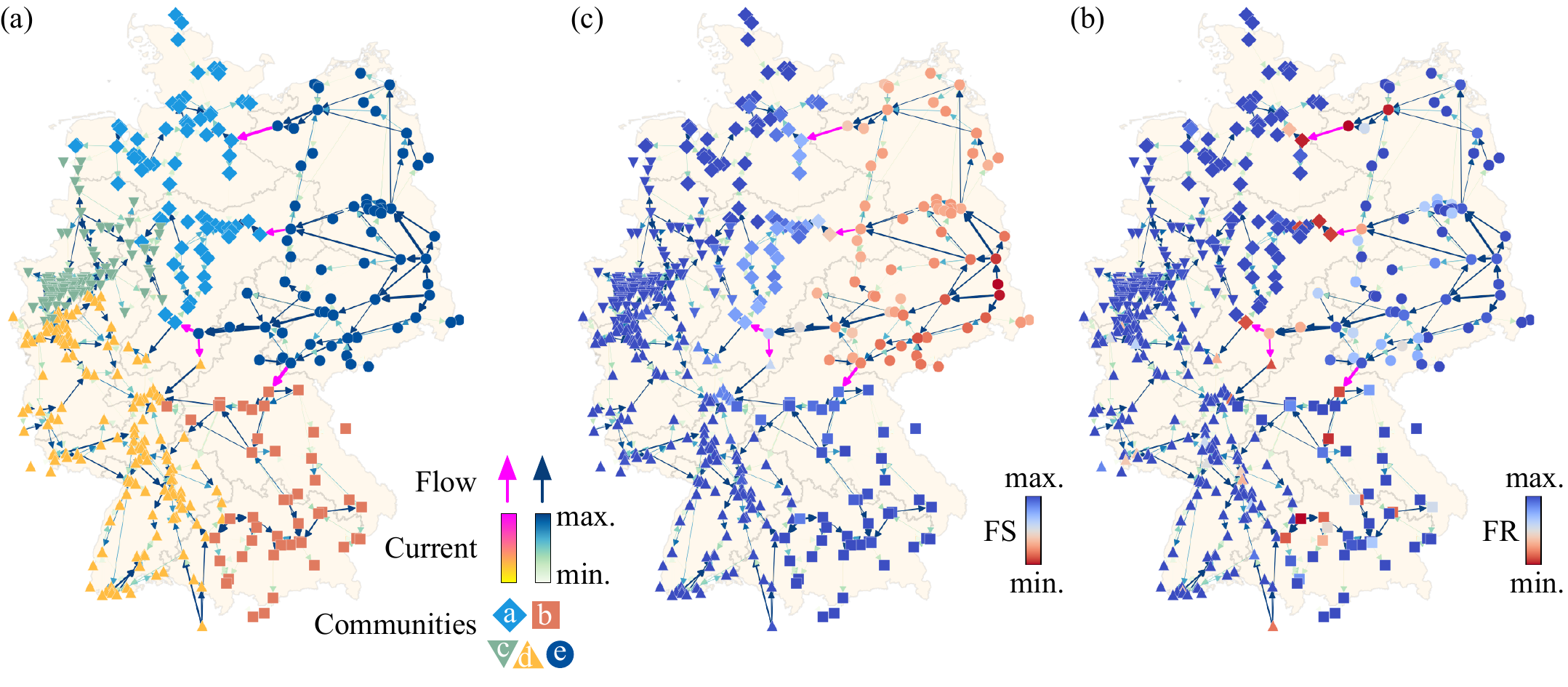}
\caption{The community structure of the German power grid and the distribution of the single-node FS and FR of nodes. (a) The community structure of the German power grid at $\gamma=0.2$ with five communities reveal that the community `e' only sends power (pink arrows) to the neighbor communities. (b) The nodes with low single-node FS are located in the community `e'. The pink out-flow links are the bridges connecting the community `e' to the others. (c) The nodes located at the bridges feature lower single-node FR than the others. Color bar scale: the node color changes linearly between minimum $0.6$ and maximum $1.0$ for FS; minimum $0.6$, maximum $1.0$ for FR.
}
\label{fig_5}
\end{figure*}

The community partition of the German power grid is shown in Fig.~\ref{fig_5}(a).
Each shape of the nodes represent each community, and the arrows indicate the direction of current flow of the links at steady state with its width and color for the amount.
It is interesting to see that the boundary of the community `e' (navy-color and round-shape) features only out-flow current (highlighted by the pink arrows) from the community.
It means that the community `e' is a power-producer group in the power grid by only supplying electricity to the other part of the grid.
The other communities both receive and send power from or to neighbor communities. 
However, only the community `e' in the Eastern Germany plays a single role as a producer group.

The community structure of the German power grid is strongly related to the distribution pattern of the single-node FS and FR.
For example, the nodes that have low single-node FS are gathering in the community `e' in the Eastern Germany [Fig.~\ref{fig_5}(b)].
Notice that the community `e' has only out-flow current (pink arrows) from it.
It seems like that the nodes in a power-producer group (that is the community `e' in this case) are likely to have low functional secureness.
In fact, it agrees well with what we found in the six-node motif in Fig.~\ref{fig_2}. 
In the motif, the only power supplier, N1, showed extraordinarily lower single-node FS than the others.
In the German power grid, not a single node but a group of nodes is the net power supplier in the grid, and the nodes in the group have lower single-node FS than the others.
It implies that it is not only a node of which centralized power generation makes the functionality unstable but also a group of nodes in the community level.

It is worth to recognize that the boundary between communities is where the nodes with low single-node FR are.
For instance, in Fig.~\ref{fig_5}c, the nodes with low single-node FR are connected to the out-flow links (pink arrows) of the community `e'.
Previously, we found that the ``gate-keeper" nodes---the nodes that are at the bridges where excessive power flows---show low basin stability~\cite{Kim:2019cq}.
Considering that the community `e' is connected to the other communities by only four connection links, and the gate-keeper effect could also be valid in the community level as we see from the low-single-node-FR nodes in Fig.~\ref{fig_5}(c).
Indeed, it is intuitively understandable that the disturbance from the gate-keeper nodes can be amplified due to the excessive power flowing through them influencing many nodes surrounding it.
Therefore, the boundary between communities can be the `death valley' of the functional robustness in power grids.
The results of German power grid showed the influence of community structure to the functional scureness and robustness. 
We validate the findings from a clustered model network in the following section.
  
\subsection{\label{sec:model}Benchmark network models}

In this section, we validate the modular effect that we observed in the German power grid by reproducing the same phenomenon in a synthetic benchmark network.
Benchmark networks are modular networks that are often used to test community detection algorithms~\cite{Lancichinetti:2008js}.
To create modular benchmark networks in this study, we utilize the protocol proposed in Ref.~\cite{Kim:2019em} that enables us to tune the number of communities and the link density within a community as well as between communities.
First, we generate three Erd\H{o}s-R{\'e}nyi (ER) random subgraphs~\cite{ERgraph} that have 20, 30, and 50 nodes in each one with average degree of 3.
Then, we select 10\% of the nodes in each subgraph and rewire one of their links to a node in the other subgraphs.
All selection processes are conducted uniformly at random.
Through this process, we can merge the three subgraphs into a modular network in which each subgraph corresponds to a community.
Each node is either a power producer $P_i=1$ or a consumer $P_i=-1$, and the number of producers and consumers are equal so that $\sum_iP_i=0$.

\begin{figure*}[h]
\hfill\includegraphics[width=1\textwidth]{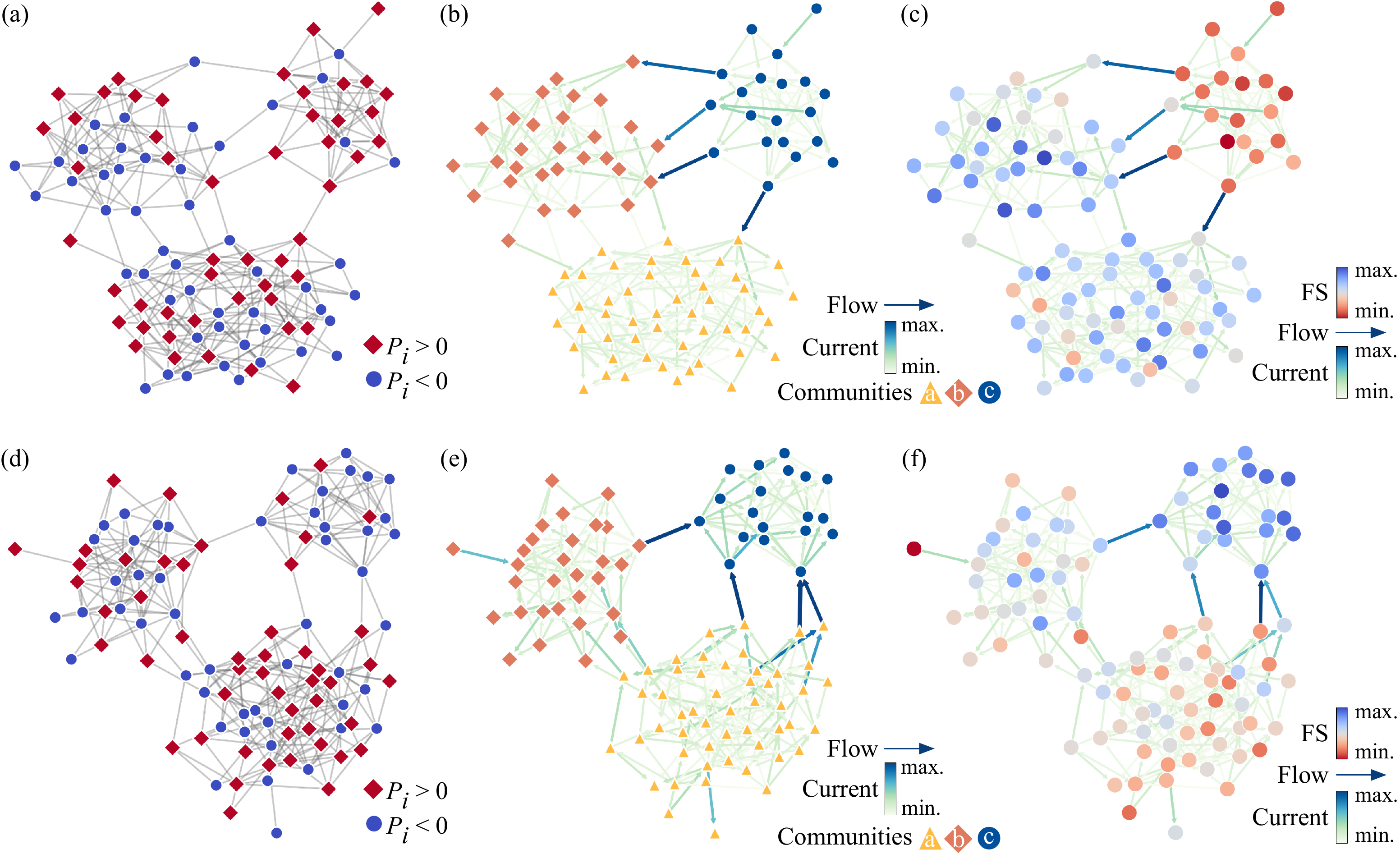}
\caption{(a) A benchmark network consists of power producers (red diamonds) and consumers (blue rounds). (b) Louvain algorithm detects three communities in the network, and the nodes in the community `c' (navy-color and round-shape) only supplies power to the other communities at steady state. (c) The nodes in the community `c' reveals lower single-node FS than the others, reproducing the same phenomenon found in the German power grid. The node color changes linearly between minimum $0.109$ and maximum $0.160$ for FS. (d) Another benchmark network also has three communities, (e) but the community `c' is now solely taking power from the network at steady state. (e) The nodes in the sole-power-consumer community have higher single-node FS than the others. The node color changes linearly between minimum $0.106$ and maximum $0.158$ for FS.
}
\label{fig_6}
\end{figure*}

We prepare two benchmark networks to test not only the case of a power-producer group as German power grid but also a power-consumer group for comparison.
For the former case, we select 15, 12, and 23 power-producer nodes uniformly at random [red diamond nodes in Fig.~\ref{fig_6}(a)] from the 20-, 30-, and 50-node-size communities, respectively.
At this configuration, due to the excessive number of power producers in the community `c' (navy-color and round-shape), the power solely flow out from the community `c' at steady state [Fig.~\ref{fig_6}(b)].
Namely, the community `c' is a power-producer group as the eastern region of German power grid.
Measuring the single-node FS values by perturbing $E=500$ times for each node, we find that the distribution pattern of the single-node FS is well reproduced in the benchmark network as the form that we observed in the German power-grid case: the nodes in the power-producer group have relatively lower single-node FS than the others [Fig.~\ref{fig_6}(c)]. 

We also examine the opposite case, a power-consumer group in a modular structure.
From a newly created benchmark network having three communities (20, 30, and 50 nodes in each one), we select 15, 12, and 23 power-consumer nodes uniformly at random [Fig.~\ref{fig_6}(d)].
In this network at steady state, the nodes in the community `c' (navy-color and round-shape) is a power-consumer group to which current flows from the other communities [Fig.~\ref{fig_6}(e)].
Contrary to the previous case, the nodes in the power-consumer group have higher single-node FS [Fig.~\ref{fig_6}(f)] meaning that they are functionally more secure. 

From the benchmark networks, we confirm that the unidirectional current flow from or to a community in a modular power grid affects the functional secureness of the associated nodes in the community.
Specifically, when a community in a power grid is a solely power provider (consumer) with only outward (inward) current flow, the nodes in the community have low (high) functional secureness. 
For the FR patterns, however, it is required to expand its size enough to form heterogeneous `influence zone' for each node. 
In our benchmark model, the network is rather small so that the distribution pattern of the single-node FR of gate-keeper nodes is not prominent. 
The relationship between the size of the network and the appearance of the `death valley' can be an interesting further research topic. 

\section{\label{sec:discussion}Summary and discussions}

In this paper, we have presented practical measures to estimate the dynamical characteristics of power-grid nodes.
The measures consider two different aspects of synchronization resilience to the rated frequency upon perturbations:
how a node can be influenced by the others, which we call functional secureness; how a node can influence to the others, functional robustness.
We have also taken into account partial synchronization in which some power-grid nodes are still synchronized, but the others are not.
Perturbations can be given to a single node at a time, multiple nodes simultaneously, or a set of particular nodes.
Through this approach, we can better understand the operational resilience of power grids in various scenarios such as power system failures or smart grid system configurations.

We have demonstrated the usefulness of functional secureness and functional robustness by a six-node motif and an IEEE test grid with 24 buses for a single-node and multi-node random perturbations.
Specifically, by single-node measures with a six-node motif we have seen that these measures can reveal the operational characteristics of power-grid nodes in terms of rated frequency. 
From the stability landscape, we have found that there exist a critical value of coupling strength where the effective interaction between nodes begins.
With the IEEE test grid we have shown that the response of the power-grid nodes against multiple perturbations can be quite different from what we can see with multi-node basin stability.
For example, the nodes in the IEEE test grid are separated into two groups depending on their stability under global disturbance.
The nodes in a group lost their synchrony, while the others were able to recover the rated frequency even when all nodes were perturbed.
The dynamical characteristics were not identifiable by the other conventional measures so that functional secureness and functional robustness can complement the investigation on the synchronization dynamics of power grids.

As a case study, we have estimated single-node functional secureness and robustness of German power-grid nodes.
Analyzing the community structure of the power grid and the current flow at steady state, we have found that the unidirectional current flow between communities affects the operational resilience of the power-grid nodes.
In particular, the eastern region of Germany formed a community of which electric power flows out to the other communities through the inter-community links. 
We have found that the nodes in the power-producing community are more vulnerable based on single-node FS.
By single-node FR, we have also observed that the nodes that are between the power-producing community and the other communities are less resistive to the perturbations.
It means that the modular structure of power grids and the consequent uni-directional current flow can affect the synchronization stability of power-grid nodes.
We have reproduced the results with two benchmark networks that include a power-producer group or a power-consumer group, respectively, 
and confirmed the phenomenal concomitance of the out-flow (in-flow) current of a community and low (high) functional secureness.

What we have observed from our simulations and results are the consequence of dynamic interaction between synchronous oscillators.
Therefore, the results are strongly related to the physical nature of power grids. 
For example, the large amount of out-flow of the power-producer community shown in German power grid means that the phase difference between the nodes of each end at the corresponding links---the gate-keeper nodes---are larger than the others. 
In addition, each of the gate-keeper nodes belongs to each community being connected to many other nodes as the definition of a community: the dense clustered connection between nodes. 
These situations help the gate-keeper nodes to be influential since its phase change (either directly caused by the perturbation on phase or indirectly as the consequence of the perturbation on frequency) can affect many nodes in its community. 
Moreover, the disturbance can easily be delivered to the other group of nodes because the gate-keeper nodes always have at least a connection to another community. 
This can trigger an avalanche of disturbance to multiple communities resulting in a large size cascade captured by low functional robustness values.

Through this paper, we have introduced the novel measures and revealed the relationship between the modular structure and synchronization stability of power grids. 
We emphasize that the mesoscale structure of power grids can affect the stability of the systems, which can be a valuable point to investigate some unique phenomena of power grids such as the power-grid islanding.
Further research is required to deepen our understanding on the fundamental mechanism of the mesoscale synchronization dynamics, e.g., the temporal propagation of disturbances in a modular structure, the influence of the community formation such as the density of inter-community links, and a coarse-graining protocol to simplify a modular power grid to an effective dynamical backbone for the accelerated simulation of the synchronization dynamics.
We hope that our measures can be a flare to initiate the relevant researches on such intriguing research questions on this subject.

\section*{Acknowledgement}
In this study HK was supported by the Chilean Council of Scientific and Technological Research, CONICYT, through the grant FONDECYT N.11190096.

\end{document}